\documentclass[prd]{revtex4}

\usepackage{bm}

\def\ba{{\bm a}}
\def\bb{{\bm b}}
\def\bh{{\bm h}}
\def\bk{{\bm k}}
\def\bl{{\bm l}}
\def\bn{{\bm n}}
\def\bv{{\bm v}}
\def\bw{{\bm w}}

\def\bx{{\bm x}}
\def\bxp{{\bm x'}}
\def\bxpp{{\bm x''}}
\def\by{{\bm y}}
\def\bD{{\bm D}}
\def\bN{{\bm N}}
\def\bR{{\bm R}}
\def\bS{{\bm S}}
\def\bW{{\bm W}}

\def\al{\alpha}
\def\l{\lambda}
\def\Ga{\Gamma}
\def\Omv{\Omega(x_{A}, x_{B})}
\def\Om{\Omega}
\def\bom{{\bm \omega}}
\def\bna{{\bm \nabla}}

\def\lb{\label}

\def\be{\begin{equation}}
\def\ee{\end{equation}}
\def\bea{\begin{eqnarray}}
\def\eea{\end{eqnarray}}

\begin{document}

\title{Time transfer and frequency shift to the order 1/\textit{c}$^{\bm 4}$ \\
in the field of an axisymmetric rotating body}

\author{Bernard Linet}
\email{linet@celfi.phys.univ-tours.fr}
\affiliation{Laboratoire de Math\'ematiques et Physique Th\'eorique, 
CNRS/UMR 6083, Universit\'e Fran\c{c}ois Rabelais, F-37200 Tours, France}
 
\author{Pierre Teyssandier}
\email{Pierre.Teyssandier@obspm.fr}
\affiliation{D\'epartement Syst\`emes de R\'ef\'erence Temps et Espace,
CNRS/UMR 8630, \\
Observatoire de Paris, 61 avenue de l'Observatoire, F-75014 Paris, France}

\date{\today}

\begin{abstract}

Within the weak-field, post-Newtonian approximation of 
the metric theories of gravity, we determine the one-way time transfer 
up to the order $1/c^4$, the unperturbed term being of order $1/c$, and
the frequency shift up to the order $1/c^4$.
We adapt the method of the world-function developed by Synge to the Nordvedt-Will 
PPN formalism. We get an 
integral expression for the world-function up to the order $1/c^3$ and
we apply this result to the field of an isolated, axisymmetric rotating body. 
We give a new procedure enabling to calculate the influence of the
mass and spin multipole moments of the body on the time transfer and the 
frequency shift up to the order $1/c^4$. We obtain explicit formulas for 
the contributions of 
the mass, of the quadrupole moment and of the intrinsic angular
momentum. In the case where the only PPN parameters different from zero are
$\beta$ and $\gamma$, we deduce from these results the complete expression 
of the frequency shift up to the order $1/c^4$. We briefly discuss the 
influence of the quadrupole moment and of the rotation of the Earth on the 
frequency shifts in the ACES mission. 

\end{abstract}


\maketitle

\section{Introduction}

Owing to recent progress in absolute frequency measurements of some optical transitions with a 
femtosecond laser, it seems possible to achieve in a near future atomic clocks having a 
time-keeping accuracy of the order of $10^{-18}$ in fractional frequency \cite{hol}. 
Since a reduced gravity would significantly
increase clock performances, it is envisaged to install such clocks on board 
artificial satellites and to compare them with terrestrial clocks by
exchange of electromagnetic signals. Already, a spatial experiment like 
ESA's Atomic Clock Ensemble in Space (ACES) mission \cite{sal,spa} is planned 
for 2006, the purpose being to obtain an accuracy of order $10^{-16}$
in fractional frequency.

At a level of uncertainty about $10^{-18}$, a fully relativistic 
treatment of time/frequency transfers must be performed up to the order 
$1/c^4$ \footnote{Let us prevent a possible confusion. When we talk about calculating
terms of order $1/c^4$ in the time transfer, we mean terms of absolute order $1/c^4$.
As a consequence, these terms are of order $1/c^3$ relative to the leading term, which is 
a distance divided by $c$. For the frequency shifts, there is no ambiguity since
we are calculating only frequency ratios.}. As far as we know, the corresponding calculations
have not been carried out. For the time transfer, the main relativistic correction 
of order $1/c^3$ is the well-known Shapiro time delay \cite{sha}. Other corrections 
due to the quadrupole moment and to the intrinsic angular momentum have been studied 
by several authors \cite{kli1}. Gravitational corrections of order $1/c^2$ in the 
frequency transfers were theoretically
determined and experimentally checked a long time ago \cite{ves}. These corrections are now 
commonplace in the Global Positioning System. The relativistic theory of the
frequency transfers have been recently extended up to the terms of order
$1/c^3$ \cite{bla1}, justifying the results previously given in \cite{ashby} 
without any detail. However, it must be pointed out that, in \cite{bla1}, 
some terms of order $1/c^3$ due to the quadrupole moment $J_2$ of the Earth
are bounded without any explicit calculation.  
Furthermore, the time/frequency transfers have been calculated within the
limited framework of general relativity, which prevents from 
discussing new tests of gravitational theories. 

The present paper is a first step towards a general theory of the time and frequency 
transfers including all the terms of order $1/c^4$, within the post-Newtonian approximation 
of any metric theory of gravity. Using the Nordtvedt-Will PPN formalism \cite{will}, we bring the 
complete determination of these effects in the field of an isolated, axisymmetric 
rotating body, the gravitational field being assumed stationary. Of course, modelling 
a mission in the vicinity of the Earth at a level of accuracy about $10^{-18}$ will 
require to add the effects due to the tidal gravitational field induced by the Sun and the Moon.

We assume that the photons ensuring the transfers
follow null geodesics. The problems that we have to tackle come down to the 
following ones, relative to a couple of points $x_A=(ct_A,\bx_A)$ and 
$x_B=(ct_B,\bx_B)$ connected by a null geodesic: 

{\em i}) to calculate the (coordinate) time transfer $t_B - t_A$ as 
a function of $(\bx_A, \bx_B)$;
 
{\em ii}) to determine the vectors tangent to the null geodesic at $x_A$ and 
$x_B$. Solving this second problem is indeed  indispensable to 
calculate the frequency shift between $x_A$ and $x_B$.   

The method generally employed to study the questions related to 
the propagation of light in a gravitational field is based on the solution of the 
null geodesic equations (see, e.g., \cite{kli1, kop1, kop2, kli3, kop3} 
for investigations in the linearized, weak-field limit of general relativity).
However, the theory of the world-function developed by Synge \cite{syn} 
presents the great advantage to spare the trouble of integrating
the geodesic equations. Once the world-function is determined, it is possible
to solve straightforwardly the two above-mentioned problems. This method is particularly
elegant for the stationary, axisymmetric field and we apply it in the present
paper. We find a new procedure enabling to determine the 
influence of the mass and spin multipole moments of the body. Explicit
calculations are given for the contributions of the mass, of the quadrupole moment
and of the intrinsic angular momentum of the rotating body. 

The paper is organized as follows. In Sec. II, the relevant properties
of the world-function $\Om(x_A, x_B)$ are recalled and the general
expression of this function in the post-Newtonian limit of any metric theory
is given. The corresponding expression of the time transfer $t_B-t_A$ 
is derived up to the order $1/c^4$. In Sec. III, we determine the expression
of $\Omega (x_A,x_B)$ and of $t_B-t_A$ within the ten-parameter PN
formalism of Nordtvedt and Will. Then, in Sec. IV, we focus our attention on the case of an
isolated, axisymmetric rotating body. We show that it is possible to determine 
the contributions of the mass and spin multipole moments by straightforward 
differentiations of a single function. Retaining only the terms due to the 
mass $M$, to the quadrupole moment $J_2$ and to the intrinsic angular momentum
$\bS$ of the rotating body, we obtain explicit expressions for the time 
transfer up to the order $1/c^4$ and for the tangent vectors at $x_A$ and 
$x_B$ up to the order $1/c^3$. In Sec. V, the frequency shift is developped up
to the order $1/c^4$ in the case where $\beta$ and $\gamma$ are the only non 
vanishing PPN parameters. We find detailed expressions for the contributions 
of $J_2$ and $\bS$ and we discuss the possible influence of these terms in 
the ACES mission. We give our conclusions in Sec. VI.

In this paper, $G$ is the Newtonian gravitational constant and $c$ is the 
speed of light in a vacuum. The Lorentzian metric of space-time is denoted 
by $g$. The signature adopted for $g$ is $ (+ - - -)$. We suppose that the 
space-time is covered by one global coordinate system 
$(x^{\mu})=(x^0,\bx )$, where $x^0=ct$, $t$ being a time coordinate, and
$\bx =(x^i)$, the $x^i$ being quasi Cartesian coordinates. We assume that 
the curves of equations $x^{i} = const.$ are timelike, which means that 
$g_{00} > 0$ anywhere. We employ the vector 
notation $\ba$ in order to denote either  $(a^1, a^2, a^3) = (a^i)$ or
$(a_1, a_2, a_3) = (a_i)$. 
Considering two such quantities $\ba$ and $\bb$ 
with for instance $\ba = (a^i)$, we use $\ba \cdot \bb$ to denote 
$a^i b^i$ if $\bb= (b^i)$ or $a^i b_i$ if $\bb = (b_i)$ 
(the Einstein convention on the repeated indices is used). The quantity 
$\mid \! \ba \! \mid$ stands for the ordinary Euclidean norm of $\ba$.

\section{The world-function and its post-Newtonian limit}

\subsection{Definition and fundamental properties}

Consider two points $x_A$ and $x_B$ in a space-time with a given metric
$g_{\mu \nu}$ and assume that $x_A$ and $x_B$ are connected by a unique
geodesic path $\Gamma$. Throughout this paper, $\lambda$ denotes the unique affine parameter 
along $\Gamma$ which fulfills the boundary conditions $\lambda_A=0$ and 
$\lambda_B=1$. The so-called world-function of space-time \cite{syn} is the 
two-point function $\Omega (x_A,x_B)$ defined by
\be \lb{1} 
\Om (x_A,x_B)=\frac{1}{2}\int_{0}^{1}g_{\mu \nu}(x^{\alpha}(\lambda ))
\frac{dx^{\mu}}{d\lambda}\frac{dx^{\nu}}{d\lambda}d\lambda \, , 
\ee
the integral being taken along $\Gamma$. It is easily seen that 
$\Om (x_A,x_B)=\varepsilon [ s_{AB}]^2/2$, where $s_{AB}$ is the
geodesic distance between $x_A$ and $x_B$ and $\varepsilon =1,0,-1$ for
timelike, null and spacelike geodesics, respectively.
It results from definition (\ref{1}) that the world-function 
$\Om (x_A,x_B)$ is unchanged if we perform any admissible coordinate 
transformation. 

The utility of the world-function for our purpose comes from the following properties \cite{syn}. 

{\em i}) Two points $x_A$ and $x_B$ are linked by a light ray if and 
only if the condition 
\be \lb{2}
\Om (x_A,x_B)=0 
\ee
is fulfilled. Thus, $\Om (x_A,x)=0$ is the equation of the light cone ${\cal C}(x_{A})$ at $x_A$. This 
fundamental property shows that if $\Om (x_A,x_B)$ is known, it is possible to determine the 
travel time $t_B - t_A$ of a photon connecting two points $x_A$ and $x_B$ as a function of 
$t_{A}$, $\bx_{A}$ and $\bx_{B}$. It must be pointed out, however, that solving the equation 
$\Om(ct_A, \bx_{A}, ct_B, \bx_{B}) = 0$ for $t_B$ 
yields two distinct solutions $t_B^{+}$ and $t_B^{-}$ since the timelike 
curve $ x^{i} = x_B^{i}$ cuts the light cone ${\cal C}(x_{A})$ at two points $x_B^{+}$ and $x_B^{-}$, 
$x_B^{+}$ being in the future of $x_B^{-}$. In the present paper, we always regard $x_A$ as the 
point of emission of the photon and $x_B$ as the point of reception, and we are concerned only 
with the determination of $t_{B}^{+} - t_{A}$ as a function of $t_{A}$, $\bx_A$ and $\bx_B$. We put
\be \lb{2a}
t_{B}^{+} - t_A = {\cal T}(t_A, \bx_{A}, \bx_{B}) \, ,
\ee
and we call ${\cal T}(t_A, \bx_{A}, \bx_{B})$ the (coordinate) time transfer function. 
Of course, it is also possible to introduce an other time transfer function 
giving $t_{B}^{+} - t_A $ as a function of the instant of reception $t_{B}^{+}$ and of $\bx_A$, $\bx_B$, 
but we do not use it here. 

{\em ii}) The vectors $(dx^{\alpha}/d\l)_A$ and $(dx^{\alpha}/d\l)_B$ tangent to the 
geodesic $\Gamma$ respectively at $x_{A}$ and $x_{B}$ are given by
\be \lb{3}
\left( g_{\alpha \beta}\frac{dx^{\beta}}{d\lambda}\right)_A=
-\frac{\partial \Omega}{\partial x_{A}^{\alpha}} \, , \quad
\left( g_{\alpha \beta}\frac{dx^{\beta}}{d\lambda}\right)_B=
\frac{\partial \Omega}{\partial x_{B}^{\alpha}} \, . 
\ee
As a consequence, if $\Omega (x_A,x_B)$ is explicitly known, the determination of these vectors 
does not require the integration of the differential equations of the geodesic. Let us note that 
it can be proved that the tangent vectors (\ref{3}) are null when (\ref{2}) holds.

Consider now a stationary space-time. In this case, we use exclusively coordinates
$(x^{\mu})$ such that the metric does not depend on $x^0$. Then, the world-function is 
a function of $x_{B}^{0} - x_{A}^{0}$, $\bx_{A}$ and $\bx_{B}$, and (\ref{2a}) reduces 
to a relation of the form 
\be \lb{4}
t_B^{+}-t_A = {\cal T}(\bx_A,\bx_B) \, .
\ee
The time transfer function ${\cal T} (\bx_{A}, \bx_{B})$ plays a central role in the 
present paper because a comparaison
between (\ref{2}) and (\ref{4}) immediately shows that the vectors 
$(l^{\mu})_A$ and $(l^{\mu})_B$ defined by their covariant components
\bea 
& & (l_0)_A=1, \quad (l_i)_A=
c\frac{\partial}{\partial x_{A}^{i}}{\cal T}(\bx_A,\bx_B) \, , \lb{6} \\
& & (l_0)_B=1, \quad (l_i)_B=
-c\frac{\partial}{\partial x_{B}^{i}}{\cal T}(\bx_A,\bx_B) \, , \lb{5}
\eea
are tangent to the ray at $x_A$ and $x_B$, respectively. It must be pointed out
that these tangent vectors correspond to an affine parameter such that
$l_0=1$ along the ray (note that such a parameter does not coincide with $\l$). Generally, 
extracting the time transfer formula
(\ref{4}) from (\ref{2}), next using (\ref{6})-(\ref{5}) will be more 
straightforward than deriving the vectors tangent at $x_A$ and $x_B$ 
from (\ref{3}), next imposing the 
constraint (\ref{2}). We shall use (\ref{6}) and (\ref{5}) in Sec. IV.

To conclude, let us emphasize that the method of the world-function works as long as 
$\Om(x_A, x_B)$ is a well-defined, single-valued
function of $x_A$ and $x_B$. This condition is satisfied in any region of space-time in 
which any points $x_A$ and $x_B$ are connected by one and only one geodesic, a feature which 
excludes the existence of conjugate points. This requirement is certainly fulfilled 
in experiments performed in the solar system and more generally for observations of stars belonging 
to our Galaxy.

\subsection{General expression of the world-function in the post-Newtonian limit}

To begin, let us assume that the metric may be written as 
\be \lb{8}
g_{\mu \nu}=\eta_{\mu \nu} + h_{\mu \nu}
\ee
throughout space-time, with $\eta_{\mu \nu} = diag(1, -1, -1, -1)$. Let $\Gamma_{(0)}$ be the 
straight line defined by the parametric equations 
$x^{\al} = x_{(0)}^{\al}(\l)$, with
\be \lb{9}
x_{(0)}^{\alpha}(\lambda) = (x_{B}^{\alpha}-x_{A}^{\alpha}) \lambda +
x_{A}^{\alpha} \, , \quad 0 \leq \l \leq 1 \, .
\ee
With this definition, the parametric equations of the geodesic $\Gamma$ 
connecting $x_A$ and $x_B$ may be written in the form
\be \lb{9a}
x^{\al}(\l) = x_{(0)}^{\al}(\l) + X^{\al}(\l) \, , \quad 0 \leq \l \leq 1  \, ,
\ee
where the quantities $X^{\al}(\l)$ satisfy the boundary conditions
\be \lb{9b}
X^{\al}(0) = 0 \, , \quad  X^{\al}(1) = 0 \, .
\ee
Inserting (\ref{8}) and $dx^{\mu}(\l)/d\l = x_{B}^{\mu}-x_{A}^{\mu} + dX^{\mu}(\l)/d\l$ 
in (\ref{1}), then developing and noting that
\be \nonumber  
\int_{0}^{1}\eta_{\mu \nu}(x_B^{\mu} - x_A^{\mu}) \frac{dX^{\nu}}{d\l} d\l = 0
\ee
by virtue of (\ref{9b}), we find the rigorous formula
\bea \lb{9c}
\Omv & = & \Om^{(0)}(x_A,x_B) + \frac{1}{2} (x_B^{\mu} - x_A^{\mu})(x_B^{\nu} - x_A^{\nu})
\int_{0}^{1} h_{\mu \nu}(x^{\alpha}(\lambda)) d \l  \\
&  & \mbox{} + \frac{1}{2} \int_{0}^{1} \left[ g_{\mu\nu}(x^{\al}(\l)) \frac{dX^{\mu}}{d\l}
\frac{dX^{\nu}}{d\l} + 
2 (x_B^{\mu} - x_A^{\mu})h_{\mu\nu}(x^{\al}(\l)) \frac{dX^{\nu}}{d\l} \right] d\l \, , \nonumber
\eea
where the integrals are taken over $\Ga$ and $ \Om^{(0)}(x_A,x_B)$ is the 
world-function in Minkowski space-time
\be \lb{7}
\Om^{(0)}(x_A,x_B)=\frac{1}{2}\eta_{\mu \nu}(x_{B}^{\mu}-x_{A}^{\mu})
(x_{B}^{\nu}-x_{A}^{\nu}) \, .
\ee

Henceforth, we shall only consider weak gravitational fields generated by 
self-gravitating extended bodies within the slow-motion, post-Newtonian 
approximation. So, we assume that the potentials $h_{\mu\nu}$ may be 
expanded as follows 
\bea \lb{1n}
& & h_{00}=\frac{1}{c^2}h_{00}^{(2)}+\frac{1}{c^4}h_{00}^{(4)}+O(6) 
\nonumber \, , \\
& & h_{0i}=\frac{1}{c^3}h_{0i}^{(3)}+O(5) \, , \quad
h_{ij}=\frac{1}{c^2}h_{ij}^{(2)}+O(4) \, .
\eea
From these expansions and from the Euler-Lagrange equations satisfied by any geodesic $\Gamma$, namely
\be \lb{2n}
\frac{d}{d\lambda}\left( g_{\alpha \beta}\frac{dx^{\beta}}{d\lambda}
\right) =\frac{1}{2}\partial_{\alpha}h_{\mu \nu}
\frac{dx^{\mu}}{d\lambda}\frac{dx^{\nu}}{d\lambda} \, ,
\ee 
it results that $X^{\mu}(\l) = O(2)$ and that $dx^{\mu}/d\l = x_{B}^{\mu}-x_{A}^{\mu} + O(2)$. As a 
consequence, $h_{\mu\nu}(x^{\al}(\l)) = h_{\mu\nu}(x_{(0)}^{\al}(\l)) + O(4)$ and the 
third and fourth terms in the r.h.s. of Eq. (\ref{9c}) are of order $1/c^4$. These features result in an 
expression for $\Om (x_A,x_B)$ as follows
\be \lb{10}
\Om (x_A,x_B)=\Om^{(0)}(x_A,x_B)+\Om^{(PN)}(x_A,x_B) + O(4) \, ,
\ee
where
\bea \lb{4n}
& & \Omega^{(PN)}(x_A,x_B) = \frac{1}{2c^2}(x_{B}^{0}-x_{A}^{0})^2
\int_{0}^{1}h_{00}^{(2)}(x_{(0)}^{\alpha}(\lambda ))d\lambda \nonumber \\
& & \qquad \qquad \qquad \quad +\frac{1}{2c^2}(x_{B}^{i}-x_{A}^{i})
(x_{B}^{j}-x_{A}^{j})\int_{0}^{1}h_{ij}^{(2)}(x_{(0)}^{\alpha}(\lambda ))
d\lambda \\
& & \qquad \qquad \qquad \quad +\frac{1}{c^3}(x_{B}^{0}-x_{A}^{0})
(x_{B}^{i}-x_{A}^{i})\int_{0}^{1}h_{0i}^{(3)}(x_{(0)}^{\alpha}(\lambda ))
d\lambda \, , \nonumber
\eea
the integral being taken over the line $\Gamma_{(0)}$ defined by (\ref{9}).

The formulas (\ref{10}) and (\ref{4n}) yield the general expression of the
world-function up to the order $1/c^3$ within the framework of the 1 PN
approximation. We shall see in the next subsection that this approximation
is sufficient to determine the time transfer function ${\cal T}(t_A,\bx_A,\bx_B)$ up to the order $1/c^4$. 
It is worthy of note that the method used above would as well lead to the expression of the 
world-function in the linearized weak-field limit previously found by Synge \cite{syn}.
 
We shall put henceforth $\bR_{AB}=\bx_B-\bx_A$ and 
$R_{AB}=\mid \! \bR_{AB}\! \mid$. Defining the quantities 
$N^{\mu}=(x_{B}^{\mu}-x_{A}^{\mu})/R_{AB}$, (\ref{7}) and (\ref{4n}) might be 
easily rewritten with these notations.

\subsection{Time transfer at the order $1/c^4$}

Suppose that $x_B$ is the point of reception of a photon emitted at $x_A$. Taking (\ref{10}) into 
account, (\ref{2}) may be written in the form
\[ 
\Om^{(0)}(x_A,x_B)+\Om^{(PN)}(x_A,x_B)=O(4) \, ,
\]
which implies the relation
\be \lb{12}
t_{B}^{+}-t_{A} = \frac{1}{c} R_{AB} - \frac{\Omega^{(PN)}(ct_A, \bx_A , ct_B^{+}, \bx_B)}{c R_{AB}}+ O(4) \, .
\ee
Using iteratively this relation, we find for the time transfer function
\be \lb{5n}
{\cal T}(t_{A}, \bx_{A}, \bx_{B}) = \frac{1}{c} R_{AB}
-\frac{\Omega^{(PN)}(ct_A ,\bx_A , ct_A + R_{AB}, \bx_B )}{cR_{AB}} + O(5) \, .
\ee
This formula shows that the time transfer ${\cal T}(t_{A}, \bx_{A}, \bx_{B})$ can be 
explicitly calculated up to
the order $1/c^4$ when $\Omega^{(PN)}(x_A,x_B)$ is known. This fundamental
result will be exploited in the following sections.

The quantity $\Omega^{(PN)}(ct_A ,\bx_A , ct_A + R_{AB}, \bx_B )$ in (\ref{5n}) may be 
written in an integral form using (\ref{4n}), in which $R_{AB}$ and $R_{AB}\l + ct_A$ are 
substituted for $x_B^{0} - x_A^{0}$ and for $x_{(0)}^{0}(\l)$, respectively. Hence
\be \lb{14}
{\cal T}(t_{A}, \bx_{A}, \bx_{B}) = \frac{1}{c}R_{AB}\left\{ 1-\frac{1}{2c^2}\int_{0}^{1}
\left[ h_{00}^{(2)}(z^{\al}(\l)) + h_{ij}^{(2)}(z^{\al}(\l))N^i N^j 
+ \frac{2}{c}h_{0i}^{(3)}(z^{\al}(\l))N^i \right] d\lambda \right\} +O(5) \, ,
\ee
the integrals being taken over the line defined by the parametric equations 
$x^{\al} = z^{\al}(\l)$, where 
\be \lb{13}
z^0(\lambda )=R_{AB}\lambda + ct_A \, , \quad
z^i(\lambda )=R_{AB}N^i\lambda +x_{A}^{i} \, , \quad 0\leq \lambda \leq 1 \, .
\ee
It must be noted that the line defined by (\ref{13}) is the null geodesic of Minkowski 
metric from $x_A$, the direction cosines of which are $N^{i} = (x_B^{i} - x_{A}^{i})/ R_{AB}$.

\section{World-function and time transfer within the Nordtvedt-Will PPN 
formalism}

\subsection{Metric in the 1 PN approximation}

In this section, we use the Nordvedt-Will post-Newtonian formalism involving 
ten parameters $\beta$, $\gamma$, $\xi$, $\alpha_1$, $\ldots$, $\zeta_4$
\cite{will}. We introduce slightly modified notations in order to be closed of the 
formalism recently proposed by Klioner and Soffel \cite{kli4}
as an extension of the post-Newtonian framework elaborated by Damour, Soffel and 
Xu \cite{dam} for general relativity. In particular, we denote by 
$\bv_r$ the velocity of the center of mass O relative to the universe
rest frame \footnote{This velocity is noted $\bw$ in Ref. \cite{will}.}.

Although our method is not confined to any particular assumption on the
matter, we suppose here that each source of the field is described by the
energy-momentum tensor of a perfect fluid
\[
T^{\mu \nu}=\rho c^2\left[ 1+\frac{1}{c^2}\left( \Pi +\frac{p}{\rho}\right)
\right] u^{\mu}u^{\nu}-pg^{\mu \nu} \, ,
\]
where $\rho$ is the rest mass density, $\Pi$ is the specific energy density
(ratio of internal energy density to rest mass density), $p$ is the
pressure and $u^{\mu}$ is the unit 4-velocity of the fluid. In this section
and in the following one, $\bv$ is the coordinate velocity $d\bx /dt$ of
an element of the fluid. We introduce the conserved mass density $\rho^*$
given by
\be \lb{M4}
\rho^*=\rho\sqrt{-g}u^0=\rho \left[ 1+\frac{1}{c^2}\left( \frac{1}{2}v^2
+3\gamma U \right) +O(4) \right] \, ,
\ee
where $g = \det (g_{\mu\nu})$ and $U$ is the Newtonian-like potential
\be \lb{M5}
U(x^0,\bx )=G\int \frac{\rho^*(x^0,\bxp )}{\mid \! \bx -\bxp \! \mid}
d^3\bxp\,  . 
\ee

In order to obtain a more simple form than the usual one for the potentials
$h_{0i}$, we suppose that the chosen $(x^{\mu})$ are related to a standard
post-Newtonian gauge $(\overline{x}^{\mu})$ by the transformation
\be \lb{M2}
x^0=\overline{x}^0 + \frac{1}{c^3}\left[ (1+ 2\xi +\alpha_2-\zeta_1 )
\partial_t\chi
-2\alpha_2\bv_r\cdot \bna \chi \right] , \quad x^i=\overline{x}^i \, ,
\ee
where $\chi$ is the superpotential defined by
\be \lb{M3}
\chi (x^0,\bx )=\frac{1}{2} G \int \rho^*(x^0,\bxp )\mid \! \bx -\bxp \! \mid 
d^3\bxp \, .
\ee
Moreover, we define $\widehat{\rho}$ by 
\bea \lb{M6}
\widehat{\rho} & = & \rho^* \left[ 1+\frac{1}{2}(2\gamma +1-2\xi +\alpha_3 +\zeta_1)
\frac{v^2}{c^2} +(1-2\beta +\xi +\zeta_2)\frac{U}{c^2}+(1+\zeta_3)
\frac{\Pi}{c^2}+(3\gamma -2\xi +3\zeta_4 )\frac{p}{\rho^*c^2} \right.
\nonumber \\
&   &  \left. \mbox{} \quad \quad \quad \quad \quad -\frac{1}{2}(\alpha_1 -\alpha_3)\frac{v^{2}_{r}}{c^2}
-\frac{1}{2}(\alpha_1-2\alpha_3) \frac{\bv_r \cdot \bv}{c^2} + O(4) \right]  \, .
\eea 
Then, the post-Newtonian potentials read
\bea 
& & h_{00}=-\frac{2}{c^2}w+\frac{2\beta}{c^4}w^2+\frac{2\xi}{c^4}\phi_W
+\frac{1}{c^4}(\zeta_1-2\xi )\phi_v-\frac{2\alpha_2}{c^4}
v_{r}^{i}v_{r}^{j}\partial_{ij}\chi +O(6) , \lb{M7} \\
& & \bh \equiv \{h_{0i}\} =\frac{2}{c^3}\left[ \left( \gamma +1 +\frac{1}{4}
\alpha_1 \right) \bw +\frac{1}{4}\alpha_1 w \, \bv_r \right] +O(5) , \lb{M8} \\
& & h_{ij}=-\frac{2\gamma}{c^2}w\delta_{ij} +O(4) \, , \lb{M9}
\eea
where
\bea 
& & w(x^0,\bx ) = G \int \frac{\widehat{\rho}(x^0,\bxp )}{\mid \! \bx -\bxp \! \mid }d^3\bxp
+\frac{1}{c^2}\left[ (1+ 2\xi +\alpha_2 -\zeta_1 )\partial_{tt}\chi
-2\alpha_2\bv_r \cdot \bna (\partial_t\chi ) \right] , \lb{M10} \\
& & \phi_W(x^0,\bx )=G^2\int \frac{\rho^*(x^0,\bxp )\rho^*(x^0,\bxpp )(\bx -\bxp )}
{\mid \! \bx -\bxp \! \mid^3}\cdot \left( \frac{\bxp -\bxpp}
{\mid \! \bx -\bxpp \! \mid}-\frac{\bx -\bxpp}{\mid \! \bxp -\bxpp \! \mid}
\right) d^3\bxp d^3\bxpp , \lb{M11} \\
& & \phi_v(x^0,\bx )=G\int \frac{\rho^*(x^0,\bxp )[\bv (x^0,\bxp )\cdot 
(\bx -\bxp )]^2}{\mid \! \bx -\bxp \! \mid^3}d^3\bxp , \lb{M12} \\
& & \bw (x^0,\bx )=G\int \frac{\rho^*(x^0,\bxp )\bv (x^0,\bxp )}
{\mid \! \bx -\bxp \! \mid}d^3\bxp \, . \lb{M13} 
\eea

\subsection{Determination of the world-function and of the time transfer}

For the post-Newtonian metric given by (\ref{M7})-(\ref{M13}),
it follows from (\ref{4n}) that $\Omega (x_A,x_B)$ may be written 
up to the order $1/c^3$ in the form given by Eq. (\ref{10}) with
\be \lb{25}
\Om^{(PN)}(x_A,x_B)=\Om^{(PN)}_{w}(x_A,x_B)+\Om^{(PN)}_{\bw}(x_A,x_B)
+\Om^{(PN)}_{\bv_r}(x_A,x_B) \, ,
\ee
where
\bea 
& & \Om^{(PN)}_{w}(x_A,x_B)=-\frac{1}{c^2} \left[ (x^{0}_{B}-x^{0}_{A})^2
+\gamma R_{AB}^{2}\right] \int_{0}^{1}w(x_{(0)}^{\alpha}(\lambda ))
d\lambda \, ,\lb{25a} \\
& & \Om^{(PN)}_{\bw}(x_A,x_B)=\frac{2}{c^3}\left( \gamma +1 + \frac{1}{4}\alpha_1 \right) 
(x^{0}_{B}-x^{0}_{A})\bR_{AB}\cdot 
\int_{0}^{1}\bw (x^{\alpha}_{(0)}(\lambda ))d\lambda \, , \lb{25b} \\
& & \Om^{(PN)}_{\bv_r}(x_A,x_B)=\frac{1}{2c^3}\alpha_1 (x^{0}_{B}-x^{0}_{A})
(\bR_{AB}\cdot \bv_r )\int_{0}^{1}w(x^{\alpha}_{(0)}(\lambda ))d\lambda \, ,
\lb{25c}
\eea
the integrals being calculated along the line defined by (\ref{9}).

The corresponding time transfer function is easily obtained by using (\ref{5n}) or 
(\ref{14}). We get
\bea \lb{6n}
{\cal T}(t_{A}, \bx_{A}, \bx_{B}) & = & \frac{1}{c}R_{AB} + \frac{1}{c^3}(\gamma +1)R_{AB}
\int_{0}^{1}w(z^{\al}(\l)) d\lambda  \nonumber \\
& - & \frac{2}{c^4}\bR_{AB}\cdot \left[ (\gamma +1+\frac{1}{4}\alpha_1 )
\int_{0}^{1}\bw (z^{\al}(\l)) d\lambda +\frac{1}{4} \alpha_{1}
\bv_r \int_{0}^{1}w(z^{\al}(\l)) d\lambda \right] +O(5) \, ,
\eea
the integral being evaluated along the curve defined by (\ref{13}). 

Let us emphasize that, since $w=U+O(2)$, $w$ may be replaced by the
Newtonian-like potential $U$ in expressions (\ref{25a})-(\ref{6n}).

\subsection{Case of stationary sources}

In what follows, we suppose that the gravitational field is generated by a
unique stationary source. Then, $\partial_t\chi=0$ and the potentials $w$ and 
$\bw$ do not depend on time. In this case, the integration involved in 
(\ref{25a})-(\ref{25c}) can be  performed by a method due to Buchdahl 
\cite{buc1}. Introducing the auxiliary variables $\by_A=\bx_A-\bxp$ and
$\by_B=\bx_B-\bxp$, and replacing in (\ref{9}) the parameter $\lambda$ by
$u=\lambda -1/2$, a straightforward calculation yields
\bea 
& & \int_{0}^{1}w(\bx_{(0)}(\lambda ))d\lambda =G \int \widehat{\rho}(\bxp )
F(\bxp ,\bx_A,\bx_B)d^3\bxp , \lb{28} \\
& & \int_{0}^{1}\bw (\bx_{(0)}(\lambda ))d\lambda =
G \int \rho^*(\bxp ) \bv (\bxp ) F(\bxp ,\bx_A,\bx_B)d^3\bxp \, , \lb{29}
\eea  
where the kernel function $F(\bxp ,\bx_A,\bx_B)$ has the expression
\[
F(\bxp  ,\bx_A,\bx_B)=\int_{-1/2}^{1/2}\frac{du}
{\mid \! (\by_B-\by_A)u+\frac{1}{2}(\by_B+\by_A)\! \mid} \, .
\]
Noting that $\by_B-\by_A=\bR_{AB}$, which implies that
$\mid \! \by_B-\by_A\! \mid =R_{AB}$, we find
\be \lb{19}
F(\bx ,\bx_A,\bx_B)=\frac{1}{R_{AB}}\ln \left( \frac{\mid \! \bx -\bx_A\! \mid
+\mid \! \bx -\bx_B\! \mid+R_{AB}}{\mid \! \bx -\bx_A\! \mid +
\mid \! \bx -\bx_B\! \mid -R_{AB}} \right) \, .
\ee 

Inserting (\ref{28}), (\ref{29}) and (\ref{19}) in (\ref{25a})-(\ref{25c}) and in 
(\ref{6n}) will enable one to obtain quite elegant expressions for $\Om^{(PN)}(x_A,x_B) $ 
and for ${\cal T}(\bx_{A}, \bx_{B})$, respectively.

\section{Isolated, axisymmetric rotating body}

Henceforth, we suppose that the light is propagating in the gravitational
field of an isolated, axisymmetric rotating body. The gravitational field
is assumed to be stationary. The main purpose of this section is to
determine the influence of the mass and spin multipole moments of the rotating
body on the coordinate time transfer and on the direction of light rays. 
From these results, it will be possible to obtain a relativistic modelling of 
the one-way time transfers and frequency shifts up to the order $1/c^4$ in a 
geocentric non rotating frame.

Since we treat the case of a body located very far from the other bodies
of the universe, the global coordinate system $(x^{\mu})$ used until now can
be considered as a local (i.e. geocentric) one. So, in agreement with the 
UAI/UGG Resolution B1 (2000) \cite{uai}, we shall henceforth denote by $W$ and 
$\bW$ the quantities $w$ and $\bw$ respectively defined by
(\ref{M10}) and (\ref{M13}) and we shall denote by
$G_{\mu \nu}$ the components of the metric. However, we shall continue here 
with using lower case letters for the geocentric coordinates in order to 
avoid too heavy notations.

The center of mass O of the rotating body being taken as the origin of the 
quasi Cartesian coordinates $(\bx )$, we choose the axis of symmetry as the 
$x^3$-axis. We assume that the body is slowly rotating about O$x^3$ with a 
constant angular velocity $\bom$, so that 
\be \lb{34a}
\bv (\bx )=\bom \times \bx \, .
\ee
In what follows, we put $r=\mid \! \bx \! \mid$, $r_A=\mid \! \bx_A \! \mid$
and $r_B=\mid \! \bx_A \! \mid$. We call $\theta$ the angle between $\bx$ and 
O$x^3$. We consider only the case where all points of the segment joining 
$\bx_A$ and $\bx_B$ are outside the body. We denote by $r_e$ the radius of the 
smallest sphere centered on O and containing the body (for celestial bodies, $r_e$ is the
equatorial radius). In this section, we assume the convergence of the multipole 
expansions formally derived below at any point outside the body, even if $r<r_e$.  

\subsection{Multipole developments of $W$ and $\bW$}

According to (\ref{M10}), (\ref{M13}) and (\ref{34a}), the gravitational potentials
$W$ and $\bW$ obey the equations
\be \lb{M14}
\bna^2 W=-4\pi G\widehat{\rho} \, , \quad 
\bna^2 \bW =-4\pi G\rho^* \bom \times \bx \, .
\ee
It follows from (\ref{M14}) that the potential 
$W$ is a harmonic function outside the rotating body. As a consequence, $W$ may be expanded 
in a multipole series of the form
\be \lb{31}
W(\bx )=\frac{GM}{r}\left[ 1-\sum_{n=2}^{\infty}J_n
\left( \frac{r_e}{r}\right)^n P_n(\cos \theta )\right] \, .
\ee
In this development, the $P_n$ are the Legendre polynomials
and the quantities $M$, $J_2 \ldots J_n \ldots$ correspond to the
generalized Blanchet-Damour mass multipole moments in general relativity 
\cite{bla2}.

In fact, taking into account the identity
\[
\frac{\partial^n}{\partial z^n}\left( \frac{1}{r}\right)=
\frac{(-1)^nn!}{r^{1+n}}P_n(z/r) \, ,\quad z=x^3 \, ,
\]  
it will be much more convenient for the computation of integral
(\ref{28}) to use the following expansion in a series of
derivatives of $1/r$
\be \lb{32}
W(\bx )=GM\left[ \frac{1}{r}-\sum_{n=2}^{\infty}\frac{(-1)^n}{n!}J_n
r_{e}^{n}\frac{\partial^n}{\partial z^n}\left( \frac{1}{r}\right) \right]\, .
\ee
According to (\ref{32}), the mass density $\widehat{\rho}$ can be developped in 
the multipole series
\be \lb{33}
\widehat{\rho}(\bx )=M\left[ \delta^{(3)}(\bx )-\sum_{n=2}^{\infty}\frac{(-1)^n}{n!}
J_nr_{e}^{n}\frac{\partial^n}{\partial z^n} \delta^{(3)}(\bx ) \right] \, ,
\ee
$\delta^{(3)}(\bx )$ being the Dirac distribution supported by the origin O.

Now, substituting (\ref{34a}) into (\ref{M13}) yields for the vector potential $\bW$
\be \lb{34b}
\bW (\bx ) = G \int \frac{\rho^* (\bxp )\bom \times \bxp}
{\mid \! \bx -\bxp \! \mid}d^3\bxp \, .
\ee
It is possible to show that this vector may be written as
\be \lb{34}
\bW=-\frac{1}{2}\bom \times \bna \cal V \, ,
\ee
where ${\cal V}$ is an axisymmetric function satisfying the Laplace
equation $\bna^2 {\cal V}=0$ outside the body. 
Consequently, we can expand $\cal V$ in a series of the form
\be \lb{35}
{\cal V}(\bx )=\frac{GI}{r}\left[ 1-\sum_{n=1}^{\infty}K_n
\left( \frac{r_e}{r}\right)^n P_n(\cos \theta )\right] \, ,
\ee 
where $I$ and the $K_n$ are constants. Inserting (\ref{35}) into (\ref{34}) 
and using the identity
\[
(n+1)P_n(z/r)+(z/r)P'_n(z/r)=P'_{n+1}(z/r) \, ,
\] 
we find for $\bW$ an expansion as follows
\be \lb{36}
\bW (\bx )=\frac{GI\bom \times \bx}{2r^3}\left[ 1 -\sum_{n=1}^{\infty}
K_n\left( \frac{r_e}{r}\right)^nP'_{n+1}(\cos \theta ) \right] \, ,
\ee
which coincides with a result previously obtained by one of us \cite{tey}. 
The coincidence shows that $I$ is the moment of inertia of the body about the
$z$-axis. Thus, the quantity $\bS=I\bom$ is the
intrinsic angular momentum of the rotating body. The coefficients $K_n$ are completely determined
by the density distribution $\rho^*$ and by the shape of the body \cite{tey, adler}.
Expansion  (\ref{36}) may also be written as 
\be \lb{37}
\bW (\bx )=-\frac{1}{2}G\bS \times \bna \left[ \frac{1}{r}
-\sum_{n=1}^{\infty}\frac{(-1)^n}{n!}K_nr_{e}^{n}
\frac{\partial^n}{\partial z^n}\left( \frac{1}{r}\right) \right] \, .
\ee
Consequently, the density of mass current can be developped in the
multipole series
\be \lb{38}
\rho^*(\bx )(\bom \times \bx )=-\frac{1}{2}\bS \times \bna \left[
\delta^{(3)}(\bx )-\sum_{n=1}^{\infty}\frac{(-1)^n}{n!}K_n r_{e}^{n}
\frac{\partial^n}{\partial z^n}\delta^{(3)}(\bx ) \right] \, ,
\ee 
a property which will be exploited in the following subsection.

\subsection{Multipole structure of the world-function}

The function $\Om^{(PN)}(x_A,x_B)$ is determined by 
(\ref{25})-(\ref{25c}) where $w$ and $\bw$ are respectively replaced by
$W$ and $\bW$. The integrals involved in the r.h.s. of 
(\ref{25})-(\ref{25c}) are given by (\ref{28}) and (\ref{29}). 
Substituting (\ref{33}) into (\ref{28})
and using the properties of the Dirac distribution, we obtain
\be \lb{39}
\int_{0}^{1}W\left( \bx_{(0)}(\lambda )\right) d\lambda =
GM\left[ 1-\sum_{n=2}^{\infty}\frac{1}{n!}J_nr_{e}^{n}
\frac{\partial^n}{\partial z^n}\right] F(\bx ,\bx_A,\bx_B)\bigg|_{\bx =0}\, .
\ee
Similarly, substituting (\ref{38}) into (\ref{29}), we get
\be \lb{40}
\int_{0}^{1}\bW \left(\bx_{(0)}(\lambda )\right) d\lambda =
-\frac{1}{2}G\bS \times \bna \left[1-\sum_{n=1}^{\infty}\frac{1}{n!}K_n
r_{e}^{n}\frac{\partial^n}{\partial z^n}\right] F(\bx ,\bx_A,\bx_B)
\bigg|_{\bx=0} \, .
\ee

These formulas show that the multipole expansion of
$\Om^{(PN)}(x_A,x_B)$ can be thoroughly calculated by straightforward
differentiations of the kernel function $F(\bx ,\bx_A,\bx_B)$ given by 
(\ref{19}). They constitute the essential result of the present paper, from 
which it would be possible to deduce the 
multipole expansions giving the time transfer and the frequency shift between A and B 
up to the order $1/c^4$.

In order to obtain explicit formulas, we shall only retain the
contributions due to $M$, $J_2$ and $\bS$ in the expansion yielding
$\Om^{(PN)}_W$ and $\Om^{(PN)}_{\bW}$. Then, 
denoting the unit vector along the $z$-axis by $\bk$
and noting that $\bS=S\bk$, we get for $\Om_{W}^{(1)}(x_A,x_B)$
\bea \lb{41}
\Om_{W}^{(PN)}(x_A,x_B) & = & -\frac{GM}{c^2}
\frac{(x_{B}^{0}-x_{A}^{0})^2+\gamma R_{AB}^{2}}{R_{AB}}
\ln \left( \frac{r_A+r_B+R_{AB}}{r_A+r_B-R_{AB}}\right) \nonumber \\
&   & \mbox{} +\frac{2GM}{c^2}J_2r_{e}^{2}
\frac{(x_{B}^{0}-x_{A}^{0})^2+\gamma R_{AB}^{2}}
{\left[ (r_A+r_B)^2-R_{AB}^{2}\right]^2}(r_A+r_B)
\left( \frac{\bk \cdot \bx_A}{r_A} +
\frac{\bk \cdot \bx_B}{r_B} \right)^2 \\
&   & \mbox{} -\frac{GM}{c^2}J_2r_{e}^{2}
\frac{(x_{B}^{0}-x_{A}^{0})^2+\gamma R_{AB}^{2}}{(r_A+r_B)^2-R_{AB}^{2}}
\left[ \frac{(\bk \times \bx_A )^2}{r_{A}^{3}}+\frac{(\bk \times \bx_B )^2}
{r_{B}^{3}} \right] +\cdots \nonumber
\eea
and for $\Om^{(PN)}_{\bW}(x_A,x_B)$
\be \lb{42}
\Om_{\bW}^{(PN)}(x_A,x_B)=\left( \gamma +1+\frac{1}{4}\alpha_1 \right)
\frac{2GS}{c^3}(x_{B}^{0}-x_{A}^{0})\frac{r_A+r_B}{r_Ar_B} 
\frac{\bk \cdot (\bx_A \times 
\bx_B)}{(r_A+r_B)^2-R_{AB}^{2}} +\cdots \,.
\ee
Finally, owing to the limit $\mid \! \alpha_1 \! \mid <0.02$ 
furnished in \cite{will}, we shall henceforth neglect all the
multipole contributions in $\Om_{\bv_r}^{(PN)}(x_A,x_B)$. Thus, we get
\be \lb{43}
\Om_{\bv_r}^{(PN)}(x_A,x_B)=\alpha_1 \frac{GM}{2c^3}(x_{B}^{0}-x_{A}^{0})
\frac{\bR_{AB}\cdot \bv_r}{R_{AB}}\ln \left( \frac{r_A+r_B+R_{AB}}
{r_A+r_B-R_{AB}}\right) +\cdots \, .
\ee

In this section and in the following one, the symbol $+\cdots$ stands for the 
contributions of higher multipole moments which are neglected. For the sake 
of brevity, when $+\cdots$ is used, we systematically omit to mention the 
symbol $O(n)$ which stands for the neglected post-Newtonian terms.  

\subsection{Time transfer function up to the order $1/c^4$}

Let us substitute $R_{AB}$ for $x_B^0 - x_A^0$ into (\ref{41})-(\ref{43}) and insert 
the corresponding expression of $\Om^{(PN)}$ into (\ref{5n}). We get an 
expression for the time transfer function as follows  
\be \lb{44}
{\cal T}(\bx_A,\bx_B)=\frac{1}{c}R_{AB}+{\cal T}_M(\bx_A,\bx_B)+
{\cal T}_{J_2}(\bx_A,\bx_B)+{\cal T}_{\bS}(\bx_A,\bx_B)+
{\cal T}_{\bv_r}(\bx_A,\bx_B) + \cdots \, , 
\ee
where
\bea 
{\cal T}_M(\bx_A,\bx_B) & = & (\gamma +1)\frac{GM}{c^3}
\ln \left( \frac{r_A+r_B+R_{AB}}{r_A+r_B-R_{AB}}\right) , \lb{45} \\
{\cal T}_{J_2}(\bx_A,\bx_B) & = & -(\gamma +1)\frac{GM}{c^3}\frac{J_2r_{e}^{2}R_{AB}}
{(r_A+r_B)^2-R_{AB}^{2}} 
\left[ \frac{2 (r_A+r_B)}{(r_A+r_B)^2-R_{AB}^{2}}
\left(\frac{\bk \cdot \bx_A}{r_A}+
\frac{\bk \cdot \bx_B}{r_B}\right)^2  \right. \nonumber \\
&   & \qquad \qquad \qquad \qquad \qquad \qquad \qquad \qquad \left.  
-\frac{(\bk \times \bx_A)^2}{r_{A}^{3}}
-\frac{(\bk \times \bx_B)^2}{r_{B}^{3}} 
\right] , \lb{46} \\
{\cal T}_{\bS}(\bx_A,\bx_B) & = & -\left( \gamma +1+\frac{1}{4}\alpha_1 \right)
\frac{2GS}{c^4} \frac{r_A+r_B}{r_Ar_B}
\frac{\bk \cdot (\bx_A \times \bx_B)}
{(r_A+r_B)^2-R_{AB}^{2}} , \lb{47} \\
{\cal T}_{\bv_r}(\bx_A,\bx_B) & = & - \, \alpha_1 \frac{GM}{2c^4} 
\frac{\bR_{AB}\cdot \bv_r}{R_{AB}} 
\ln \left( \frac{r_A+r_B+R_{AB}}{r_A+r_B-R_{AB}}\right) \, . \lb{48}
\eea

The time transfer is thus explicitly determined up to the order $1/c^4$.
The term of order $1/c^3$ given by (\ref{45}) is the well-known Shapiro time 
delay. Equations (\ref{46}) and (\ref{47}) extend results previously 
found for $\gamma = 1$ and $\alpha_1 = 0$ \cite{kli1}. However, our derivation is more 
straightforward and yields formulas which are more convenient to calculate the frequency 
shifts. As a final remark, it is worthy of note that ${\cal T}_M$
and ${\cal T}_{J_2}$ are symmetric in $(\bx_A,\bx_B)$, while ${\cal T}_{\bS}$
and ${\cal T}_{\bv_r}$ are antisymmetric in $(\bx_A,\bx_B)$.

\subsection{Directions of light rays at $x_A$ and $x_B$ 
up to the order $1/c^3$}

In order to determine the vectors tangent to the ray path at
$x_A$ and $x_B$, we use Eqs. (\ref{6}) and (\ref{5}) where ${\cal T}$ is
replaced by the expression given by (\ref{44}). For the sake of brevity,
we put henceforth $\bl_A =\{ (l_i)_A\}$ and $\bl_B = \{ (l_i)_B\}$. We find
\be \lb{49a} 
\bl_A(\bx_A,\bx_B)=-\bN_{AB}+\bl_M(\bx_A,\bx_B)+\bl_{J_2}(\bx_A,\bx_B)
+\bl_{\bS}(\bx_A,\bx_B)+\bl_{\bv_r}(\bx_A,\bx_B) + \cdots \, ,
\ee
\be \lb{49b} 
\bl_{B}(\bx_A,\bx_B)=-\bN_{AB}-\bl_M(\bx_B,\bx_A)-\bl_{J_2}(\bx_B,\bx_A)
+\bl_{\bS}(\bx_B,\bx_A)+\bl_{\bv_r}(\bx_B,\bx_A) + \cdots \, , 
\ee
where $\bl_M$, $\bl_{J_2}$, $\bl_{\bS}$ and $\bl_{\bv_r}$ stand for the
contributions of ${\cal T}_M$, ${\cal T}_{J_2}$, ${\cal T}_{\bS}$
and ${\cal T}_{\bv_r}$, respectively. Putting
\[
\bn_A=\frac{\bx_A}{r_A} , \quad \bn_B=\frac{\bx_B}{r_B} , \quad
\bN_{AB}=\frac{\bx_B-\bx_A}{R_{AB}} \, , 
\]
we get from (\ref{45})
\be \lb{50}
\bl_M(\bx_A,\bx_B)=-(\gamma +1)\frac{2GM}{c^2}\frac{(r_A+r_B)
\bN_{AB}+R_{AB}\bn_A}{(r_A+r_B)^2-R_{AB}^{2}} \, .
\ee
From (\ref{46}), we get
\bea \lb{51}
& & \bl_{J_2}(\bx_A,\bx_B) = (\gamma +1)\frac{GMJ_2r_{e}^{2}}
{c^2}\frac{r_A + r_B}{\left[ (r_A+r_B)^2-R_{AB}^{2}\right]^2}  \nonumber \\
& & \qquad \qquad \qquad  \; \times \left\{ \bN_{AB} 
\left[ 2(\bk\cdot \bn_A+\bk\cdot \bn_B)^2
\frac{(r_A+r_B)^2+3R_{AB}^{2}}{(r_A+r_B)^2-R_{AB}^{2}} \right. 
\right. \nonumber \\ 
& & \qquad \qquad \qquad \quad \left. -\left( \frac{1-(\bk\cdot \bn_A)^2}{r_A}+
\frac{1-(\bk\cdot \bn_B)^2}{r_B}
\right)\frac{ (r_A+r_B)^2+R_{AB}^{2}}{r_A + r_B} \right] \nonumber \\
& & \qquad \qquad \qquad \quad  +2\bn_A \frac{R_{AB}}{r_A + r_B}\left[ (\bk\cdot \bn_A+\bk\cdot \bn_B)^2
\frac{3(r_A+r_B)^2+R_{AB}^{2}}{(r_A+r_B)^2-R_{AB}^{2}} \right.  \\ 
& & \qquad \qquad \qquad \quad -\frac{1}{2}\left( 1-3(\bk\cdot \bn_A)^2\right) 
\frac{(3r_A+r_B)(r_A+r_B)-R_{AB}^{2}}{r_{A}^{2}}  \nonumber \\
& & \qquad \qquad \qquad \quad \left. +(r_A+r_B)\left( \frac{2(\bk\cdot \bn_A)(\bk\cdot \bn_B)}{r_A}
-\frac{1-(\bk\cdot \bn_B)^2}{r_B}\right) \right] \nonumber \\
& & \qquad \qquad \qquad \quad \left. -4\bk \frac{R_{AB}}{r_A}\left[(\bk\cdot \bn_A)
\frac{ (3r_A+r_B)(r_A+r_B)-R_{AB}^{2}}{2r_A (r_A + r_B)}
 +(\bk \cdot \bn_B )\right] \right\} \, . \nonumber
\eea 
From (\ref{47}) and (\ref{48}), we derive the other contributions which
are of order $1/c^3$
\bea 
& & \bl_{\bS}(\bx_A,\bx_B)=\left( \gamma + 1 + \frac{1}{4}\alpha_1\right) 
\frac{2GS}{c^3}\frac{r_A+r_B}{r_A[(r_A+r_B)^2-R_{AB}^{2}]} 
\bigg\{ \bk \times \bn_B  \nonumber \\
& & \qquad \qquad \qquad  +\frac{2r_Ar_B \, \bk\cdot (\bn_A\times \bn_B)}
{(r_A+r_B)^2-R_{AB}^{2}}
\left[ \frac{(3r_A+r_B)(r_A+r_B)-R_{AB}^{2}}{2r_A(r_A+r_B)}\bn_A+\bn_B
\right] \bigg\} \, , \lb{52} \\
& & \bl_{\bv_r}(\bx_A,\bx_B)=\alpha_1\frac{GM}{c^3}
\left[ \frac{\bv_r-(\bv_r\cdot \bN_{AB})
\bN_{AB}}{2R_{AB}}\ln \left( \frac{r_A+r_B+R_{AB}}{r_A+r_B-R_{AB}}
\right) \right. \nonumber \\
& & \qquad \qquad \qquad \qquad \qquad \qquad \qquad \qquad \quad  \left. +(\bv_r\cdot \bN_{AB})\frac{(r_A+r_B)\bN_{AB}+R_{AB}\bn_A}
{(r_A+r_B)^2-R_{AB}^{2}}\right] \, . \lb{53}
\eea

We note that the mass and the quadrupole moment yield contributions of
order $1/c^2$, while the intrinsic angular momentum and the velocity
relative to the universe rest frame yield contributions of order $1/c^3$.

\subsection{Sagnac terms in the time transfer function}

In an experiment like ACES, recording the time of emission $t_A$ will be more
practical than recording the time of reception $t_B$. So, it will be very
convenient to form the expression of the time transfer ${\cal T}(\bx_A , \bx_B )$ from
$\bx_A(t_A)$ to $\bx_B(t_B^{+})$ in terms of the position of the receiver B
at the time of emission $t_A$. For any quantity $Q_B(t)$ defined along
the world line of the station B, let us put $\widetilde{Q}_B=Q(t_A)$. Thus we
may write $\widetilde{\bx}_B(t_A)$, $\widetilde{r}_B(t_A)$, $\widetilde{\bv}_B(t_A)$, 
$\widetilde{v}_B = \mid \! \widetilde{\bv}_B \! \mid$, etc.

Now, let us introduce the instantaneous coordinate distance 
$\bD_{AB}=\widetilde{\bx}_B-\bx_A$ and its norm $D_{AB}$. Since we want 
to know $t_B-t_A$ up to the order $1/c^4$, we can use the Taylor expansion of
$\bR_{AB}$
$$
\bR_{AB}=\bD_{AB}+(t_B-t_A)\widetilde{\bv}_B+\frac{1}{2}(t_B-t_A)^2 \,
\widetilde{\ba}_B+\frac{1}{6}(t_B-t_A)^3 \, \widetilde{\bb}_B+\cdots \, ,
$$
where $\ba_B$ is the acceleration of B and $\bb_B=d\ba_B/dt$. Using 
iteratively this expansion together with (\ref{44}), we get
\bea \lb{55}
& & {\cal T}(\bx_A,\bx_B)={\cal T}(\bx_A,\widetilde{\bx}_B)+ 
\frac{1}{c^2}\bD_{AB} \cdot \widetilde{\bv}_B 
+\frac{1}{2c^3}D_{AB} \left[ \frac{(\bD_{AB}\cdot \widetilde{\bv}_B)^2}
{D_{AB}^{2}}+\widetilde{v}_{B}^{2}+\bD_{AB}\cdot \widetilde{\ba}_B\right] \nonumber \\
& & \qquad \qquad \qquad +\frac{1}{c^4}\left[ \left( \bD_{AB}\cdot 
\widetilde{\bv}_B\right) \left( \widetilde{v}_{B}^{2}
+\bD_{AB}\cdot \widetilde{\ba}_B \right) +\frac{1}{2}D_{AB}^{2} \left(
\widetilde{\bv}_B\cdot \widetilde{\ba}_B+\frac{1}{3}\bD_{AB}\cdot \widetilde{\bb}_B \right)
\right]  \\
& & \qquad \qquad \qquad +\frac{1}{c}\frac{\bD_{AB}}{D_{AB}}\cdot 
\widetilde{\bv}_B \left[ {\cal T}_M(\bx_A,\widetilde{\bx}_B)+{\cal T}_{J_2}(\bx_A,\widetilde{\bx}_B) \right] 
\nonumber \\
& & \qquad \qquad \qquad +\frac{1}{c^2}D_{AB}\widetilde{\bv}_B \cdot \left[ \bl_M(\widetilde{\bx}_B,\bx_A)
+\bl_{J_2}(\widetilde{\bx}_B,\bx_A)\right] + \cdots \, , \nonumber
\eea
where ${\cal T}(\bx_A,\widetilde{\bx}_B)$ is obtained by substituting 
$\widetilde{\bx}_B$, $\widetilde{r}_B$ and $\bD_{AB}$ respectively for $\bx_B$, $r_B$ 
and $\bR_{AB}$ into the time transfer function defined by 
(\ref{44})-(\ref{48}). This expression extends the previous formula \cite{bla1} to the 
next order $1/c^4$. The second, the third
and the fourth terms in (\ref{55}) represent pure Sagnac terms
of order $1/c^2$, $1/c^3$ and $1/c^4$, respectively. The fifth and the sixth terms are 
contributions of the gravitational field mixed with the coordinate velocity of the 
receiving station. Since these last two terms are of order $1/c^4$, they may be 
calculated for the arguments $(\bx_B,\bx_A)$ (note the order of the arguments 
in $\bl_M$ and $\bl_{J_2}$).

\section{Frequency shift in the field of an axisymmetric rotating body}

\subsection{General formulas}

Consider a clock ${\cal O}_A$ on A and a clock ${\cal O}_B$ on B 
delivering respectively the proper frequencies $f_A$ and $f_B$ and
suppose that ${\cal O}_A$ is sending photons to ${\cal O}_B$. The
one-way frequency transfer from ${\cal O}_A$ and ${\cal O}_B$ is
characterized by the ratio $f_A/f_B$ which may be written as
$f_A/f_B=(f_A/\nu_A)(\nu_A/\nu_B)(\nu_B/f_B)$ where $\nu_A$ is the
proper frequency of the photon as measured on A at the instant of emission
and $\nu_B$ is the proper frequency of the same photon as measured on B 
at the instant of receipt. The ratios $f_A/\nu_A$ and $f_B/\nu_B$ are obtained 
by local measurements performed on A and B, respectively \cite{bla1}. So, in the present
paper, we are concerned only with the theoretical determination of
$\nu_A/\nu_B$. This ratio is given by the well-known relation
\be \lb{56}
\frac{\nu_A}{\nu_B}=\frac{u_{A}^{\mu}(l_{\mu})_A}
{u_{B}^{\mu}(l_{\mu})_B}
\ee
where $u_{A}^{\mu}=(dx^{\mu}/ds)_A$ and 
$u_{B}^{\mu}=(dx^{\mu}/ds)_B$ are respectively the unit 4-velocity of 
the clock ${\cal O}_A$ and of the clock ${\cal O}_B$, and $(l_{\mu})_A$ and 
$(l_{\mu})_B$ are the null tangent vectors at the point of emission $x_A$ and at 
the point of reception $x_B$, respectively.

Let us denote by $\bv_A=(d\bx/dt)_A$ and $\bv_B=(d\bx/dt)_B$ the
coordinate velocities of the clocks on A and B, respectively. Since
the gravitational field is assumed to be stationary, the formula 
(\ref{56}) giving the frequency shift between $x_A$ and $x_B$ may be written as
\be \lb{57}
\frac{\nu_A}{\nu_B}=\frac{u_{A}^{0}}{u_{B}^{0}}\times
\frac{q_A}{q_B} \, ,\quad
q_A=1+\frac{1}{c}\bl_A\cdot \bv_A \, , \quad 
q_B=1+\frac{1}{c}\bl_B\cdot \bv_B \, ,
\ee
where $\bl_A$ and $\bl_B$ are
the quantities respectively defined by (\ref{6}) and (\ref{5}). 

It is possible to calculate the ratio $q_A/q_B$ up to the order $1/c^4$
from our results in Sec. IV since $\bl_A$ and $\bl_B$ are given up to
the order $1/c^3$ respectively by (\ref{49a}) and (\ref{49b}). Denoting by
$\bl^{(n)}/c^n$ the $O(n)$ terms in $\bl$, $q_A/q_B$ may be expanded as
\bea \lb{59}
& & \frac{q_A}{q_B}=1-\frac{1}{c}\frac{\bN_{AB}\cdot (\bv_A-\bv_B)}
{1 \displaystyle - \bN_{AB}\cdot \frac{\bv_B}{c}} 
+\frac{1}{c^3}\left[ \bl_{A}^{(2)}\cdot \bv_A-\bl_{B}^{(2)}\cdot \bv_B 
\right] +\frac{1}{c^4}\left[ \bl_{A}^{(3)}\cdot \bv_A-\bl_{B}^{(3)}
\cdot \bv_B \right] \nonumber \\
& & \qquad \; +\frac{1}{c^4}\bN_{AB}\cdot \left[ \left( \bl_{B}^{(2)}\cdot \bv_B\right)
(\bv_A-2\bv_B)+\left( \bl_{A}^{(2)}\cdot \bv_A 
\right) \bv_B \right]+ O(5)  \, . 
\eea 

In order to be consistent with this expansion, we have to perform the
calculation of $u_{A}^{0}/u_{B}^{0}$ at the same level of approximation.
For a clock delivering a proper time $\tau$, $1/u^0$ is the ratio of the
proper time $d\tau$ to the coordinate time $dt$. To reach the suitable
accuracy, it is therefore necessary to take into account the terms
of order $1/c^4$ in $g_{00}$. For the sake of simplicity, we shall henceforth 
confine ourselves to the fully conservative metric theories of gravity 
without preferred location effects, in which all the PPN parameters vanish 
except $\beta$ and $\gamma$. Since the gravitational field is assumed to be
stationary, the chosen coordinate system is then a standard 
post-Newtonian gauge and the metric reduces to its usual form 
\be \lb{M15}
G_{00}=1-\frac{2}{c^2}W+\frac{2\beta}{c^4}W^2+O(6) , \quad
\{G_{0i}\}=\frac{2(\gamma +1)}{c^3}\bW +O(5) , \quad
G_{ij}=-\left( 1+\frac{2\gamma}{c^2}W \right) \delta_{ij}+O(4) \, ,
\ee
where $W$ given by (\ref{M10}) reduces to
\be \lb{M18}
W(\bx )=U(\bx )+\frac{G}{c^2}\int \frac{\rho^*(\bxp )}{\mid \! \bx -\bxp \! \mid}
\left[ \left( \gamma +\frac{1}{2}\right) v^2+(1-2\beta )U+\Pi
+3\gamma \frac{p}{\rho^*}\right] d^3 \bxp \, ,
\ee
and $\bW$ is given by (\ref{34b}). As a consequence, for a clock moving with 
the coordinate velocity $\bv$, the quantity $1/u^0$ is given by the formula
\be \lb{61}
\frac{1}{u^0}\equiv \frac{d\tau}{dt}=1-\frac{1}{c^2}\left( W+
\frac{1}{2}v^2\right) 
+\frac{1}{c^4}\left[ \left( \beta -\frac{1}{2} \right) W^2
-\left( \gamma +\frac{1}{2}\right) Wv^2-\frac{1}{8}v^4
+2(\gamma +1)\bW\cdot \bv \right] +O(6) \, ,
\ee
from which it is easily deduced that
\bea \lb{65}
\frac{u_{A}^{0}}{u_{B}^{0}} = 1 & + & \frac{1}{c^2}\left( W_A-W_B+
\frac{1}{2}v_{A}^{2}-\frac{1}{2}v_{B}^{2}\right) \nonumber \\ 
& + & \frac{1}{c^4}\bigg\{ (\gamma +1)(W_Av_{A}^{2}-W_Bv_{B}^{2})
+\frac{1}{2}(W_A-W_B)\left[ W_A-W_B+v_{A}^{2}-v_{B}^{2}+2(1-\beta )
(W_A+W_B) \right] \\
&   & \mbox{} \qquad \qquad \qquad \qquad + \frac{3}{8}v_{A}^{4}-\frac{1}{4}v_{A}^{2}v_{B}^2
-\frac{1}{8}v_{B}^{4}-2(\gamma +1)(\bW_A\cdot \bv_A -\bW_B\cdot \bv_B) \bigg\}
+O(6) \nonumber \, .
\eea
  
It follows from (\ref{59}) and (\ref{65}) that the frequency shift
$\delta \nu /\nu$ is given by
\be \lb{66}
\frac{\delta \nu}{\nu}\equiv
\frac{\nu_A}{\nu_B}-1=\left( \frac{\delta \nu}{\nu}\right)_c +
\left( \frac{\delta \nu}{\nu}\right)_g \, ,
\ee
where $(\delta \nu /\nu )_c$ is the special-relativistic Doppler effect
\bea \lb{67}
& & \left( \frac{\delta \nu}{\nu}\right)_c = 
-\frac{1}{c}\bN_{AB}\cdot (\bv_A-\bv_B) 
+\frac{1}{c^2}\left[ \frac{1}{2}v_{A}^{2}-\frac{1}{2}v_{B}^{2}-
\left( \bN_{AB}\cdot (\bv_A-\bv_B)\right) \left( \bN_{AB}\cdot \bv_B \right)
\right] \nonumber \\
& & \qquad \qquad -\frac{1}{c^3}\left[ \left( \bN_{AB}\cdot (\bv_A-\bv_B)\right)
\left( \frac{1}{2}v_{A}^{2}-\frac{1}{2}v_{B}^{2}+
\left( \bN_{AB}\cdot \bv_B\right)^2 \right) \right] \\
& & \qquad \qquad +\frac{1}{c^4}\left[ \frac{3}{8}v_{A}^{4}-\frac{1}{4}v_{A}^{2}v_{B}^{2}
-\frac{1}{8}v_{B}^{4} 
-\left( \bN_{AB}\cdot (\bv_A-\bv_B)\right) \left( \bN_{AB}\cdot 
\bv_B \right) \left( \frac{1}{2}v_{A}^{2}-\frac{1}{2}v_{B}^{2}
+\left( \bN_{AB}\cdot \bv_B \right)^2 \right) \right] +O(5) \nonumber
\eea  
and $(\delta \nu )/\nu )_g$ contains all the contribution of the
gravitational field, eventually mixed with kinetic terms
\bea \lb{68}
& & \left( \frac{\delta \nu}{\nu}\right)_g=\frac{1}{c^2}(W_A-W_B) 
-\frac{1}{c^3}\left[ (W_A-W_B)\left( \bN_{AB}\cdot (\bv_A-\bv_B)\right) -
\bl_{A}^{(2)}\cdot \bv_A+\bl_{B}^{(2)}\cdot \bv_B \right] \nonumber \\
& & \qquad \qquad +\frac{1}{c^4}\left\{ (\gamma +1)(W_Av_{A}^{2}-W_Bv_{B}^{2})
+\frac{1}{2}(W_A-W_B)\left[ W_A-W_B+2(1-\beta )(W_A+W_B)+v_{A}^{2}-v_{B}^{2} 
\right. \right. \\
& & \qquad \qquad \left. \left. -2\left( \bN_{AB}\cdot (\bv_A-\bv_B) \right) 
\left( \bN_{AB}\cdot \bv_B \right) \right]
+ \bN_{AB}\cdot \left[ \left( \bl_{B}^{(2)}\cdot \bv_B \right) 
(\bv_A-2\bv_B)+\left( \bl_{A}^{(2)}\cdot \bv_A \right) \bv_B \right] 
\right.  \nonumber \\ 
& & \qquad \qquad \left. +\left( \bl_{A}^{(3)}-2(\gamma +1)
\bW_A \right)\cdot \bv_A
-\left( \bl_{B}^{(3)}-2(\gamma +1)\bW_B \right) \cdot \bv_B \right\}
+O(5) \nonumber \, .
\eea

It must be emphasized that the formulas (\ref{61}) and (\ref{65}) are valid within the PPN
framework without adding special assumption, provided that $\beta$ and
$\gamma$ are the only non-vanishing post-Newtonian parameters. On the
other hand, (\ref{68}) is valid only for stationary gravitational fields.
In the case of an axisymmetric rotating body, we shall obtain an
approximate expression of the frequency shift by inserting the following
developments in (\ref{68}), yielded by (\ref{49a})-(\ref{53}):   
\bea
& & \bl_{A}^{(2)}/c^2=\bl_M(\bx_A,\bx_B)+\bl_{J_2}(\bx_A,\bx_B)+ \cdots 
\, , \quad 
\bl^{(3)}_{A}/c^3= \bl_{\bS}(\bx_A,\bx_B)+ \cdots \, , \nonumber \\
& & \bl_{B}^{(2)}/c^2=-\bl_M(\bx_B,\bx_A)-\bl_{J_2}(\bx_B,\bx_A) 
+\cdots \, , \quad
\bl^{(3)}_{B}/c^3=\bl_{\bS}(\bx_B,\bx_A)+\cdots \, , \nonumber 
\eea
the function $\bl_{\bS}$ being now given by (\ref{52}) written with
$\alpha_1=0$. Let us recall that the symbol $+ \cdots$ stands for the contributions 
of the higher multipole moments which are neglected.

\subsection{Application in the vicinity of the Earth}

In order to perform numerical estimates of the frequency shifts in the 
vicinity of the Earth, we suppose now that A is on board the International
Space Station (ISS) orbiting at the altitude $H=400$ km and that B is a
terrestrial station. It will be the case for the ACES mission. We use 
$r_B = 6.37 \times 10^6$ m and $r_A - r_B =400$ km. For 
the velocity of ISS, we take $v_A = 7.7 \times 10^3$ m/s and for the 
terrestrial station, we have $v_B \leq 465$ m/s. The other useful parameters 
concerning the Earth are: $G M = 3.986 \times 10^{14}$ m$^3$/s$^2$, 
$r_e = 6.378 \times 10^6$ m, $J_2 = 1.083 \times 10^{-3}$; for 
$n\geq 3$, the multipole moments $J_n$ are in the order of $10^{-6}$. With 
these values, we get $W_B/c^2 \approx G M /c^2 r_B = 6.95 \times 10^{-10}$ and 
$W_A /c^2 \approx  G M /c^2 r_A = 6.54 \times 10^{-10}$. From these data, it is 
easy to deduce the following upper bounds:
$\mid \! \bN_{AB} \cdot \bv_A /c \! \mid \leq  2.6 \times 10^{-5}$ 
for the satellite, 
$\mid \! \bN_{AB}\cdot \bv_B /c \! \mid \leq 1.6 \times 10^{-6}$ 
for the ground station and 
$\mid \! \bN_{AB}\cdot (\bv_A-\bv_B)/c \! \mid \leq 2.76 \times 10^{-5}$
for the first-order Doppler term.

Our purpose is to obtain correct estimates of the effects
in (\ref{68}) with are greater than or equal to $10^{-18}$ for an axisymmetric
model of the Earth. At this level of approximation, it is not sufficient
to take into account the $J_2$-terms in $(W_A-W_B)/c^2$. First, the
higher-multipole moments $J_3$, $J_4$, $\ldots$ yield contribution of order
$10^{-15}$ in $W_A/c^2$. Second, owing to the irregularities in the
distribution of masses, the expansion of the geopotential in a series of
spherical harmonics is probably not convergent at the surface of the
Earth. For these reasons, we do not expand $(W_A-W_B)/c^2$ in (\ref{68}).

However, for the higher-order terms in (\ref{68}), we can apply the 
explicit formulas obtained in the previous section. Indeed, since the
difference between the geoid and the reference ellipsoid is less than
$100$ m, $W_B/c^2$ may be written as \cite{wol}
\[
\frac{1}{c^2}W_B=\frac{GM}{c^2r_B}+\frac{GMr_{e}^{2}J_2}{2c^2 r_{B}^{3}}
(1-3\cos^2\theta )+\frac{1}{c^2}\triangle W_B \, ,
\]
where the residual term $\triangle W_B/c^2$ is such that 
$\mid \! \triangle W_B/c^2 \! \mid \leq 10^{-14}$.
At a level of experimental uncertainty about $10^{-18}$, this inequality 
allows to retain only the contributions due to $M$,
$J_2$ and $\bS$ in the terms of orders $1/c^3$ and $1/c^4$. As a
consequence, the formula (\ref{68}) reduces to  
\bea \lb{0665}
& & \left(\frac{\delta \nu}{\nu} \right)_g=\frac{1}{c^2}(W_A-W_B)+
\frac{1}{c^3}\left( \frac{\delta \nu}{\nu} \right)_{M}^{(3)} + 
\frac{1}{c^3}\left( \frac{\delta \nu}{\nu} \right)_{J_2}^{(3)} 
+\cdots \nonumber \\ 
& & \qquad \qquad +\frac{1}{c^4}\left( \frac{\delta \nu}{\nu} \right)_{M}^{(4)}+ 
\frac{1}{c^4}\left( \frac{\delta \nu}{\nu} \right)_{\bS}^{(4)}+\cdots \, , 
\eea
where the different terms involved in the r.h.s. are separately explicited and
discussed in what follows.

Using the identity $(r_A+r_B)^2-R_{AB}^{2}=2r_Ar_B(1+\bn_A\cdot \bn_B)$,
it may be seen that $(\delta \nu /\nu )_{M}^{(3)}$ is given by
\bea \lb{06652}
& & \left( \frac{\delta \nu}{\nu} \right)_{M}^{(3)}=-\frac{GM(r_A+r_B)}{r_Ar_B}
\left[ \left( \frac{\gamma +1}{1+\bn_A\cdot \bn_B}-\frac{r_A-r_B}{r_A+r_B}
\right) \bN_{AB}\cdot (\bv_A-\bv_B) \right. \nonumber \\
& & \qquad \qquad \; \left. +(\gamma +1)\frac{R_{AB}}{r_A+r_B}\frac{\bn_A\cdot \bv_A+\bn_B
\cdot \bv_B}{1+\bn_A\cdot \bn_B}\right] \, .
\eea
The contribution of this third-order term is bounded by $5\times 10^{-14}$
for $\gamma =1$, in accordance with a previous analysis \cite{bla1}.

\subsection{Influence of the quadrupole moment at the order $1/c^3$}

Defining the quantity $K_{AB}$ by 
\[
K_{AB}=\frac{(r_A- r_B)^2}{r_A r_B}\, ,
\]
it is easily deduced from 
(\ref{51}) and (\ref{68}) that the term 
$\left(\delta \nu /\nu \right)_{J_{2}}^{(3)}$ in (\ref{0665}) is given by
\bea 
\left(\frac{\delta \nu}{\nu} \right)_{J_2}^{(3)} & = & \frac{G M}{2r_e} J_{2} 
\left( \bN_{AB} \cdot (\bv_A - \bv_B ) \right) \left[
\left( \frac{r_e}{r_A} \right)^3 \left[ 3 (\bk \cdot \bn_A)^2 - 1 \right] - 
\left(\frac{r_e}{r_B} \right)^3 
\left[ 3 (\bk \cdot \bn_B)^2 - 1 \right] \right] \nonumber \\ 
&  & \mbox{} + \, \frac{\gamma +1}{2}\, 
\frac{GM \, J_2 r_{e}^{2}(r_A + r_B)}{r_A^2  r_B^2} \frac{1}
{(1 + \bn_A \cdot \bn_B)^2}  \nonumber \\   
&  & \mbox{} \times \bigg\{ (\bN_{AB}\cdot (\bv_A - \bv_B))
\left[ (\bk \cdot \bn_A + \bk \cdot \bn_B )^2 \, 
\frac{5 - 3 \bn_A \cdot \bn_B + 2K_{AB}}{1 + \bn_A \cdot \bn_B} \right. \nonumber \\
& & \qquad \qquad \left. - \left( 1 - \frac{r_A (\bk \cdot \bn_B)^2 + 
r_B (\bk \cdot \bn_A)^2}{r_A +r_B} \right) 
(3 - \bn_A \cdot \bn_B + K_{AB}) \right]   \nonumber \\
& & \qquad \qquad + \frac{R_{AB}}{r_A + r_B} (\bn_A \cdot \bv_A + \bn_B \cdot \bv_B ) 
(\bk \cdot \bn_A + \bk \cdot \bn_B )^2 \,   
\frac{7 - \bn_A \cdot \bn_B +2 K_{AB} }{1 + \bn_A \cdot \bn_B}  \nonumber \\ 
& & \qquad \qquad - \frac{R_{AB}}{r_A}( \bn_A \cdot \bv_A ) \left[1 - 3(\bk \cdot \bn_A)^2 \right] 
\frac{r_A +  r_B (2 + \bn_A \cdot \bn_B)}{r_A + r_B} \lb{0741} \\
& & \qquad \qquad - \frac{R_{AB}}{r_B} ( \bn_B \cdot \bv_B ) \left[1 - 3(\bk \cdot \bn_B)^2 \right]  
\frac{r_A (2 + \bn_A \cdot \bn_B ) + r_B}{r_A + r_B} \nonumber \\
& & \qquad \qquad + R_{AB} \left[ 2 \left( \frac{\bn_A \cdot \bv_A}{r_A} + \frac{\bn_B \cdot \bv_B}{r_B} \right)
(\bk \cdot \bn_A)(\bk \cdot \bn_B) \right. \nonumber \\
& &  \qquad \qquad \qquad \left. -(\bn_A \cdot \bv_A ) \frac{1-(\bk \cdot \bn_B)^2}{r_B} 
 -(\bn_B \cdot \bv_B ) \frac{1 - (\bk \cdot \bn_A)^2}{r_A} \right] \nonumber \\
& &  \qquad \qquad - 2\, \frac{R_{AB}}{r_A }( \bk \cdot \bv_A ) \left[ \bk \cdot \bn_A 
\frac{r_A + r_B(2 + \bn_A \cdot \bn_B )}{r_A + r_B}  + \bk \cdot \bn_B \right]   \nonumber \\
& &  \qquad \qquad -2 \, \frac{R_{AB}}{r_B}( \bk \cdot \bv_B ) \left[ \bk \cdot \bn_A  
+ \bk \cdot \bn_B  \frac{r_A (2 + \bn_A \cdot \bn_B ) + r_B}{r_A + r_B} \right] \bigg\} \nonumber \, . 
\eea 

One has $\mid \bv_A /c \mid = 2.6 \times 10^{-5}$, 
$\mid \bv_B /c \mid \leq 1.6 \times 10^{-6} $ and $K_{AB}=3.77 \times 10^{-3}$. A crude estimate 
can be obtained by neglecting in (\ref{0741}) the terms involving the scalar products $\bn_B \cdot \bv_B$
and $\bk \cdot \bv_B$. Since the orbit of ISS is almost circular, the scalar product 
$\bn_A \cdot \bv_A$ can also be neglected. On these assumptions, we find for $\gamma =1$
\be \lb{07414}
\bigg| \frac{1}{c^3} \left( \frac{\delta \nu}{\nu}\right)_{J_2}^{(3)} \bigg| \leq  1.3 \times 10^{-16} .
\ee 

As a consequence, it will perhaps be necessary to take into account the $O(3)$ contributions of $J_{2}$ 
in the ACES mission. This conclusion is to be compared 
with the order of magnitude given in \cite{bla1} without a detailed calculation. Of course, a better 
estimate might be found if the inclination $i = 51.6 \deg$ of the 
orbit with respect to the terrestrial 
equatorial plane and the latitude $\pi /2 - \theta_B$ of the 
ground station were taken into account.

\subsection{Frequency shifts of order $1/c^4$}

The term $(\delta \nu /\nu)_{M}^{(4)}$ in (\ref{0665}) is given by
\bea 
& & \left(\frac{\delta \nu}{\nu} \right)_{M}^{(4)} =  
(\gamma + 1)\left(\frac{GM}{r_A}v_A^2 - \frac{GM}{r_B}v_B^2 \right) 
-\frac{GM (r_A-r_B)}{2 \, r_A r_B}
(v_A^2 - v_B^2) \nonumber \\
& & \qquad \qquad + \, \frac{1}{2} \left(\frac{GM}{r_A r_B}\right)^2 
\left[(r_A-r_B )^2+2(\beta -1)(r_A^2 -r_B^2) \right]  \lb{06652a} \\
& & \qquad \qquad \; -\frac{GM(r_A+r_B)}{r_Ar_B}\left[ \left( \frac{2(\gamma +1 )}
{1+\bn_A\cdot \bn_B}-\frac{r_A-r_B}{r_A+r_B}\right) \left( \bN_{AB}
\cdot (\bv_A-\bv_B)\right) \left( \bN_{AB}\cdot \bv_B \right) 
\right. \nonumber \\
& & \qquad \qquad \; \left. +\frac{\gamma + 1}{1+\bn_A\cdot \bn_B}
\frac{R_{AB}}{r_A+r_B}
\left( (\bn_A\cdot \bv_A)\left( \bN_{AB}\cdot \bv_B\right)
-( \bN_{AB}\cdot (\bv_A-2\bv_B))(\bn_B\cdot \bv_B)\right) \right] 
\nonumber \, .
\eea
The dominant term $(\gamma + 1) GM v_A^2/r_A$ in (\ref{06652a}) induces a correction to the frequency
shift which amounts to $10^{-18}$. So, it will certainly be necessary to take 
this correction into account in experiments performed in the foreseeable future.

The terms $(\delta \nu /\nu)_{\bS}^{(4)}$ is the contribution of the 
intrinsic angular momentum to the frequency shift. 
Substituting (\ref{36}) and (\ref{52}) into (\ref{68}), it may be seen that
\be \lb{075}
\left( \frac{\delta \nu}{\nu} \right)_{\bS}^{(4)}  = 
\left({\cal F}_{\bS}  \right)_A- \left({\cal F}_{\bS} \right)_B \, ,
\ee 
where
\bea 
\left({\cal F}_{\bS} \right)_A & = &  (\gamma + 1)\, \frac{GS }{r_{A}^{2}} 
\left(1 + \frac{r_A}{r_B} \right)\bv_{A} \cdot 
\left\{ \frac{\bk \times \bn_B}{1 + \bn_A \cdot \bn_B}  
- \frac{r_B}{r_A + r_B} \, \bk \times \bn_A  \right.  \nonumber \\
          &   & \mbox{} + \left. \frac{\bk\cdot (\bn_A \times \bn_B)}{(1 + \bn_A \cdot \bn_B)^2 }
\left[  \frac{r_A +  r_B(2 + \bn_A \cdot \bn_B)}{r_A + r_B} \bn_A  + \bn_B \right] \right\}  \, , \lb{078}
\eea
\bea 
\left({\cal F}_{\bS } \right)_B & = &  (\gamma + 1) \, \frac{GS}{r_{B}^{2}} 
\left(1+ \frac{r_B}{r_A} \right) \bv_{B} \cdot 
\left\{ \frac{\bk \times \bn_A}{1 +  \bn_A \cdot \bn_B } 
- \frac{r_A}{r_A + r_B}\, \bk \times \bn_B  \right. \nonumber \\
          &   & \mbox{} - \left. \frac{\bk \cdot (\bn_A \times \bn_B)}{(1 + \bn_A \cdot \bn_B)^2 }
\left[ \bn_A +  \frac{r_A (2 + \bn_A \cdot \bn_B) + r_B}{r_A + r_B} \bn_B \right] \right\}  \, . \lb{079}
\eea

In order to make easier the discussion, it is useful to introduce the angle $\psi$ 
between $\bx_A$ and $\bx_B$ and the angle $i_p$ between the plane of the 
photon path and the equatorial plane. These angles are defined by 
\[
\cos \psi = \bn_A \cdot \bn_B \, , \quad 0\leq \psi < \pi \, , \quad
\bk\cdot (\bn_A \times \bn_B) = \sin \psi \cos i_p \, , \quad
0 \leq i_p < \pi \, .
\]
With these definitions, it is easily seen that
\[
\frac{\bk\cdot (\bn_A \times \bn_B)}{1 + \bn_A \cdot \bn_B } = 
\cos i_p \tan \frac{\psi}{2} \, .
\]
Let us apply our formulas to ISS. Due to the 
inequality $v_B / v_A \leq 6 \times 10^{-2}$, the term 
$\left({\cal F}_{\bS} \right)_B$ in (\ref{075}) may be  neglected. From (\ref{078}), 
it is easily deduced that
\[
\mid \left({\cal F}_{\bS} \right)_A \mid \leq \, (\gamma + 1)\, 
\frac{GS}{r_{A}^{2}} \,  
\left(1+ \frac{r_{A}}{ r_{B}} \right) \,  \frac{2 + 3 
\mid \! \tan  \psi /2 \! \mid }{\mid \! 1 + \cos \psi \! \mid } 
\, v_{A} \, .
\]
Assuming $0 \leq \psi \leq \pi /2$, we have 
$(2+3\mid \! \tan \psi /2 \! \mid )/ \mid \! 1+\cos \psi \! \mid \leq 5$. 
Inserting this inequality in the previous one and taking for the Earth 
$G S / c^3 r_{A}^{2} = 3.15 \times 10^{-16}$, we find 
\be \lb{084}
\bigg| \frac{1}{c^4}\left( \frac{\delta \nu}{\nu}\right)_{\bS}^{(4)} \bigg| 
\leq \, (\gamma +1)\times 10^{-19} \, . 
\ee

Thus, we get an upper bound which 
is slightly greater than the one estimated by retaining only the term 
$h_{0i}v^i/c$ in (\ref{65}). However, our formula confirms that the intrinsic 
angular momentum of the Earth will not affect the ACES experiment.

\section{Conclusion}

In this paper, we have shown that the world-function $\Om (x_A, x_B)$ 
constitutes a powerful tool for determining the coordinate time transfer and 
the frequency shift in a weak gravitational field. Our main results are
established within the Nordtvedt-Will PPN formalism. We have found the general
expression of $\Om (x_A,x_B)$ up to the order $1/c^3$. This result yields 
the expression of the time transfer function ${\cal T}(t_{A}, \bx_{A}, \bx_{B})$ 
at the order $1/c^4$. We point out that 
$\gamma$ and $\alpha_1$ are the only post-Newtonian parameters involved
in the expressions of the world-function and of the time transfer function.

We have treated in detail the case of an isolated, axisymmetric rotating body,
assuming that the gravitational field is stationary and that the body
is moving with a constant velocity $\bv_r$ relative to the universe rest 
frame. We have given a systematic procedure for calculating the terms due to
the multipole moments in the world-function $\Om (x_A,x_B)$ and in the time 
transfer function ${\cal T}(\bx_A,\bx_B)$. These terms are 
obtained by straightforward differentiations of a kernel function. 
We have explicitly derived the contributions due to the mass $M$, to the
quadrupole moment $J_2$ and to the intrinsic angular momentum $\bS$ of
the rotating body.

Restricting our attention to the case where only $\beta$ and $\gamma$ 
are different from zero, we have then determined the general expression of
the frequency shift up to the order $1/c^4$. We have obtained
the contributions of $J_2$ at the order $1/c^3$. Our method would give as 
well the quadrupole contribution at the order $1/c^4$ in case of necessity.
We have found the complete evaluation of the effect of
the intrinsic angular momentum $\bS$, which is of order $1/c^4$.
It is noteworthy that our formulas contain terms which have not been
taking into account until now.

Within the limits of our model, the formulas that we have established yield all 
the gravitational corrections to the frequency shifts up to $10^{-18}$ in the 
vicinity of the Earth. We have applied our results to the ACES mission. 
We have found that the influence of the quadrupole moment at the order $1/c^3$ is 
in the region of $10^{-16}$. For the effect of the intrinsic angular momentum, we have 
obtained an upper bound which is greater than the currently accepted estimate but which 
remains three orders of magnitude less than the expected accuracy in an experiment like ACES.
Finally, it must be noted that our results could be applied to the
two-way time/frequency transfers. In particular, the $O(3)$ contributions 
of $J_2$ to the two-way frequency transfers would probably deserve to be 
carefully calculated.


\end{document}